\documentclass[12pt,colorlinks=true,linkcolor=blue,citecolor=blue,urlcolor=blue]{article}
\makeatletter
\let\frontmatter@title@above=\relax
\makeatother
\usepackage{times}
\usepackage{geometry}
\usepackage[utf8]{inputenc}
\usepackage{enumitem,amssymb}
\usepackage{ragged2e}
\usepackage{setspace}
\usepackage{graphicx}
\usepackage{floatrow}
\usepackage{makecell}
\usepackage{color}
\usepackage{style_aastex}
\usepackage{wrapfig}
\usepackage{todonotes}

\usepackage[bf,sf,small,compact]{titlesec}
\titlespacing*{\section}{0pt}{10pt}{6pt}
\titlespacing*{\subsection}{0pt}{6pt}{3pt}
\titlespacing*{\subsubsection}{0pt}{6pt}{3pt}

\begin{document}
\pagenumbering{gobble}

\noindent {\fontsize{16}{20} \selectfont SSP 2022 White Paper}
\begin{center}
{\fontsize{24}{32}\selectfont Frequency Agile Solar Radiotelescope}
\vspace{0.3cm}

\textit{\fontsize{14}{18}\selectfont A Next-Generation Radio Telescope for Solar Physics and Space Weather}
\vspace{-0.5cm}
\end{center}
\normalsize
\begin{figure}[h]
    \centering
    \includegraphics[width=1.0\textwidth]{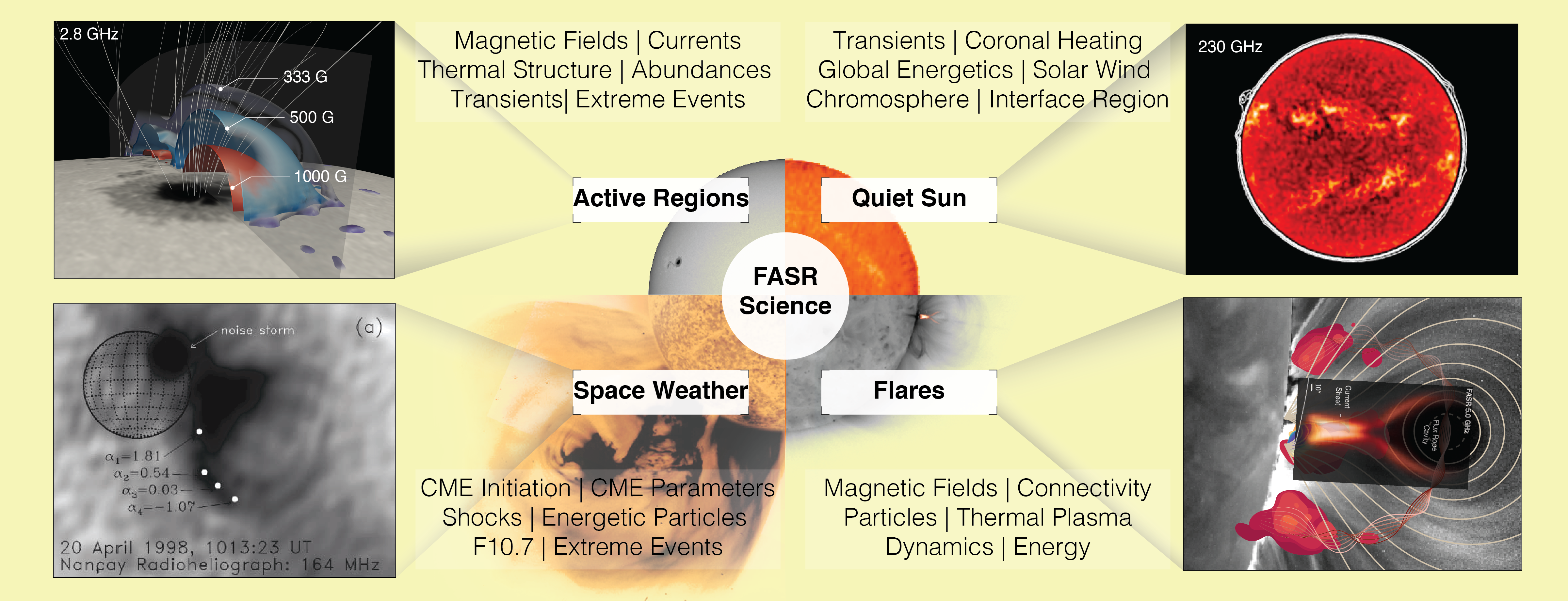}
    \label{fig:my_label}
\end{figure}
\vspace{-0.5cm}

\noindent \textbf{Principal Author:} 
Dale E. Gary, \emph{New Jersey Institute of Technology}; Email: dgary@njit.edu; Phone: (973) 642-7878; \\
\textbf{Co-authors:} 
Bin Chen, \emph{NJIT}; James F. Drake, \emph{UMD}; Gregory D. Fleishman, \emph{NJIT}; Lindsay Glesener, \emph{UM}; Pascal Saint-Hilaire, \emph{UC/Berkeley}; Stephen M. White, \emph{AFRL}; and many others on the accompanying spreadsheet.
\\
\noindent \textbf{Executive Summary:}

The {\sl Frequency Agile Solar Radiotelescope} (FASR) has been strongly endorsed as a top community priority by both Astronomy \& Astrophysics Decadal Surveys and Solar \& Space Physics Decadal Surveys in the past two decades. Although it was developed to a high state of readiness in previous years (it went through a CATE analysis and was declared “doable now”), the NSF has not had the funding mechanisms in place to fund this mid-scale program. Now it does, and the community must seize this opportunity to modernize the FASR design and build the instrument in this decade. The concept and its science potential have been abundantly proven by the pathfinding {\sl Expanded Owens Valley Solar Array} (EOVSA), which has demonstrated a small subset of FASR's key capabilities such as dynamically measuring the evolving magnetic field in eruptive flares, the temporal and spatial evolution of the electron energy distribution in flares, and the extensive coupling among dynamic components (flare, flux rope, current sheet). The FASR concept, which is orders of magnitude more powerful than EOVSA, is low-risk and extremely high reward, exploiting a fundamentally new research domain in solar and space weather physics. Utilizing {\bf dynamic broadband imaging spectropolarimetry} at radio wavelengths, with its unique sensitivity to coronal magnetic fields and to both thermal plasma and nonthermal electrons from large flares to extremely weak transients, the ground-based FASR will make synoptic measurements of the coronal magnetic field and map emissions from the chromosphere to the middle corona in 3D. 
With its high spatial, spectral, and temporal resolution, as well as its superior imaging fidelity and dynamic range, FASR is poised to provide a system-wide perspective on myriad coupled phenomena. FASR will be a highly complementary and synergistic component of solar and heliospheric observing capabilities that is critically needed to support the next generation of solar science.

\pagebreak
\section{Introduction}

\pagenumbering{arabic}
\setcounter{page}{1}

Radio observations of the Sun provide a unifying perspective among multi-wavelength observations, because of their sensitivity to both thermal plasma and nonthermal particles and their unique sensitivity to solar magnetic fields—the energy source of solar activity. The \textbf{spatially-resolved radio spectrum} provides a powerful source of diagnostic information with the potential for transformational insights into solar activity and its terrestrial impacts. 

The need to develop the necessary ground-based infrastructure and analysis tools to fully exploit radio observations of the Sun has long been recognized but is so far largely unrealized. The Frequency Agile Solar Radiotelescope (FASR) meets this need. FASR is a next-generation solar-dedicated radio telescope (a radioheliograph)
combining superior ``snapshot" imaging capability and ultra-broad frequency coverage to address a wide range of science goals.

The value of FASR to the solar, heliophysics, and space weather science communities – as well as the astronomy and astrophysics community – has been recognized by four (!) previous decadal surveys, two in Astronomy and Astrophysics \citep{2001Decadal,2010Decadal} and two in Solar and Space Physics \citep{2003Decadal, 2013Decadal}. Although the most recent Astronomy and Astrophysics survey (Astro2020; \citealt{2021Decadal}) elected not to prioritize solar ground-based facilities, it did reaffirm the need for FASR and described it as a ``missed opportunity." FASR was ranked as the number one ground-based project in the previous Solar and Space Physics decadals, while the Astro2010 decadal radio-millimeter-submillimeter panel ranked FASR second, behind HERA. Astro2010 performed a CATE analysis of FASR, described it as ``doable now” and singled it out as an exemplar of an ideal mid-scale program. While individual sections of NSF developed mid-scale infrastructure budget lines, it was only recently that it became available as an agency-wide ``big idea”. Therefore, opportunities to convert these strong recommendations into investments in FASR development have been lacking. The one outstanding exception has been an NSF MRI-R2 grant for the Expanded Owens Valley Solar Array (EOVSA), a 13-antenna pathfinder array that leveraged and validated previous FASR design work. Some important EOVSA results are discussed in the next section.

This white paper outlines steps necessary to restart the FASR initiative to bring this unique and powerful instrument to fruition.

\section{FASR Science Goals and Objectives}

FASR is a radio interferometric array designed to perform Fourier-synthesis imaging. FASR is an instrument designed and optimized for high-fidelity radio spectral analysis over the extreme range of flux density and timescales presented by the Sun. The science FASR addresses is as broad as solar physics itself, but FASR's science goals cannot be adequately addressed by non-solar-dedicated, general-purpose radio facilities. The major advances offered by FASR over previous generations of solar radio instruments are its unique combination of \textbf{ultra-wide frequency coverage, high spectral and time resolution, and superior image quality}. FASR measures the polarized brightness temperature spectrum along every line of sight to the Sun as a function of time. The concept proposed here would operate from 0.2 to 20 GHz, while yet lower frequencies (20--88 MHz) are covered by the solar-dedicated upgrade to the Long Wavelength Array at the Owens Valley Radio Observatory (OVRO-LWA), which is now in its commissioning stage. Radiation over this vast wavelength range probes the solar atmosphere from the middle chromosphere to several solar radii into the middle corona---the dynamic, magnetoactive, plasma environment in which a wealth of astrophysical and space weather processes occurs. By virtue of its broad frequency coverage, FASR will image the entire solar atmosphere multiple times per second from the chromosphere through the corona, while retaining the capability to image a selected frequency range with as little as 20 ms time resolution. FASR is sensitive to temperatures from $<10,000$~K to $>30$~MK, and nonthermal particle energies from $\sim20$ keV to $>1$ MeV.  Moreover, FASR’s panoramic view allows the solar atmosphere and the physical phenomena therein, both thermal and nonthermal, to be studied as a coupled system.

Here we summarize several main science goals of the proposed instrument while at the same time emphasizing the fundamentally new observables enabled by FASR. With its unique and innovative capabilities, FASR also has tremendous potential for new discoveries beyond those presently anticipated. 


\subsection{The Nature and Evolution of Coronal Magnetic Fields}

Quantitative knowledge of coronal magnetic fields is fundamental to understanding essentially all solar phenomena above the photosphere, including the structure and evolution of solar active regions, magnetic energy release, charged particle acceleration, flares, coronal mass ejections (CMEs), coronal heating, the solar wind and, ultimately, space weather and its impact on Earth. Characterized as the solar and space physics community’s ``dark energy” problem \citep{2004ApJ...613L.177L}, useful quantitative measurements of the coronal magnetic field have been demonstrated with breakthrough flare observations by the FASR pathfinder EOVSA \citep[e.g.][]{2018ApJ...863...83G,2020Sci...367..278F,2020NatAs...4.1140C} as well as broadband observations of solar magnetic active regions by EOVSA and the Jansky Very Large Array (JVLA). Figure~2 illustrates the use of radio observations for measuring the coronal magnetic field in active regions (ARs). See the white papers by \cite{Gary2022a} and \cite{Chen2022c} for coronal magnetic field measurements of ARs, coronal cavities, and CMEs. Such measurements are complementary to numerical extrapolations of the magnetic field distribution at the photospheric or chromospheric level \citep{2009ApJ...696.1780D}, as well as ongoing efforts at O/IR wavelengths to make measurements of above-the-limb coronal magnetic fields via the Hanle and Zeeman effects (e.g., with DKIST and COSMO; see discussion in \citealt{Gibson2021}). Coordinating with radio observations (which is already underway with EOVSA), DKIST and COSMO would enable us to derive the vector magnetic field at the places where various acceleration and transport mechanisms trigger or operate. The relation between the inferred vector magnetic field and energy spectra of accelerated particles will strongly constrain the magnetic reconnection and acceleration mechanisms in transient energy release events such as flares, CMEs, and jets \citep[e.g.][]{2021PhRvL.126m5101A}. 
See the next subsection for more discussion.

\begin{figure}[!ht] \label{fig:AR_mapping}
{\includegraphics[width=0.7\textwidth]{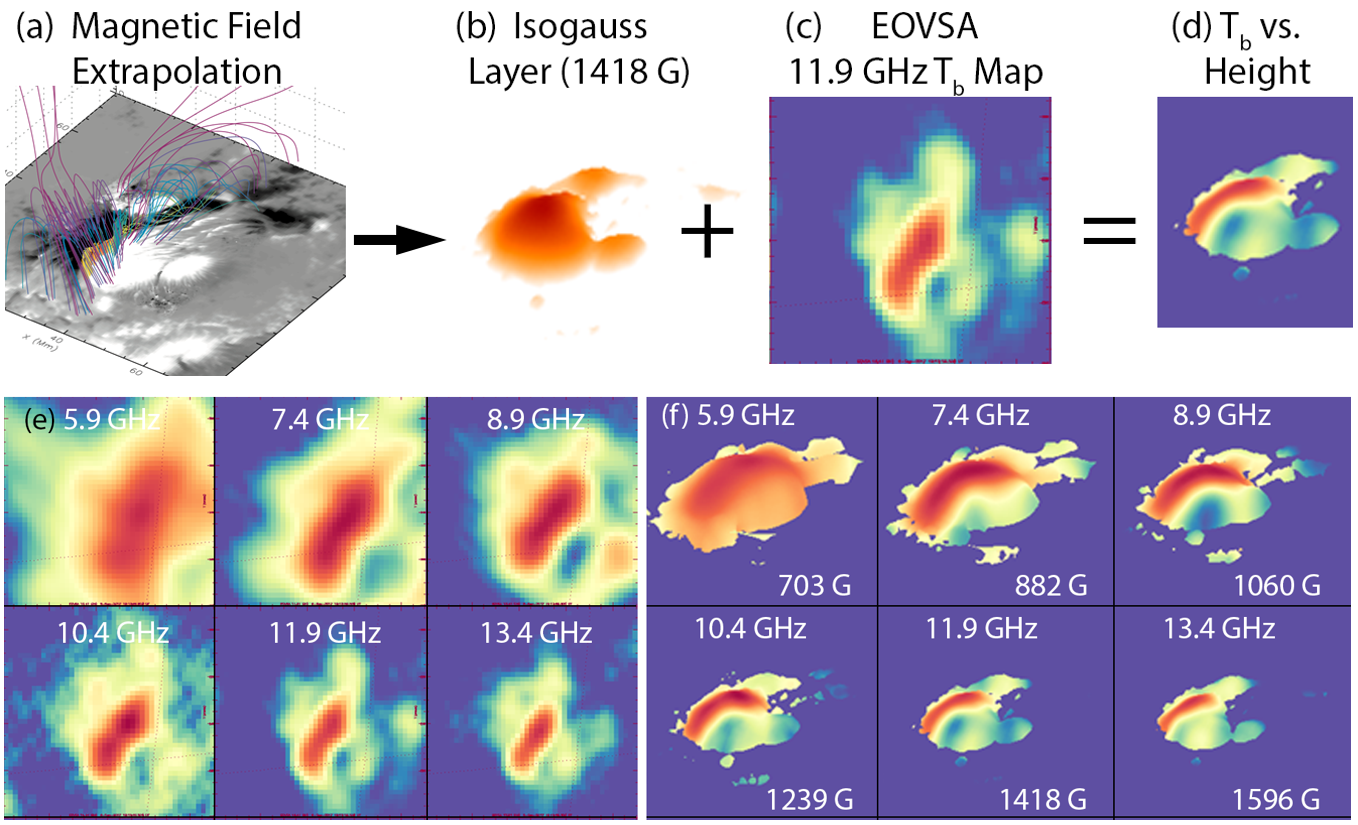}}
\caption{Illustration of the CAT-scan-like probe of active regions provided by microwave emission. (a) Perspective view of selected magnetic field lines from a NLFFF extrapolation based on the vector magnetogram of 2017-Sep-06 (gray-scale base map). (b) Perspective view of the third-harmonic isogauss layer (1418 G) obtained from the extrapolation. (c) The corresponding EOVSA 11.9 GHz brightness temperature image (peak $T_b$ is about $10^7$~K). (d) The isogauss layer ``painted” with the EOVSA brightness temperature. (e) EOVSA brightness temperature at 6 representative frequencies from 5.9-13.4 GHz, separated by 1.5 GHz, where red shades range from 7-10 MK. (f) Perspective views of isogauss surfaces at the corresponding field strengths.}
\end{figure}

\subsection{The Physics of Flares}

\begin{figure}[!ht]
\floatbox[{\capbeside\thisfloatsetup{capbesideposition={right,top},capbesidewidth=4.2cm}}]{figure}[\FBwidth]
{\includegraphics[width=0.7\textwidth]{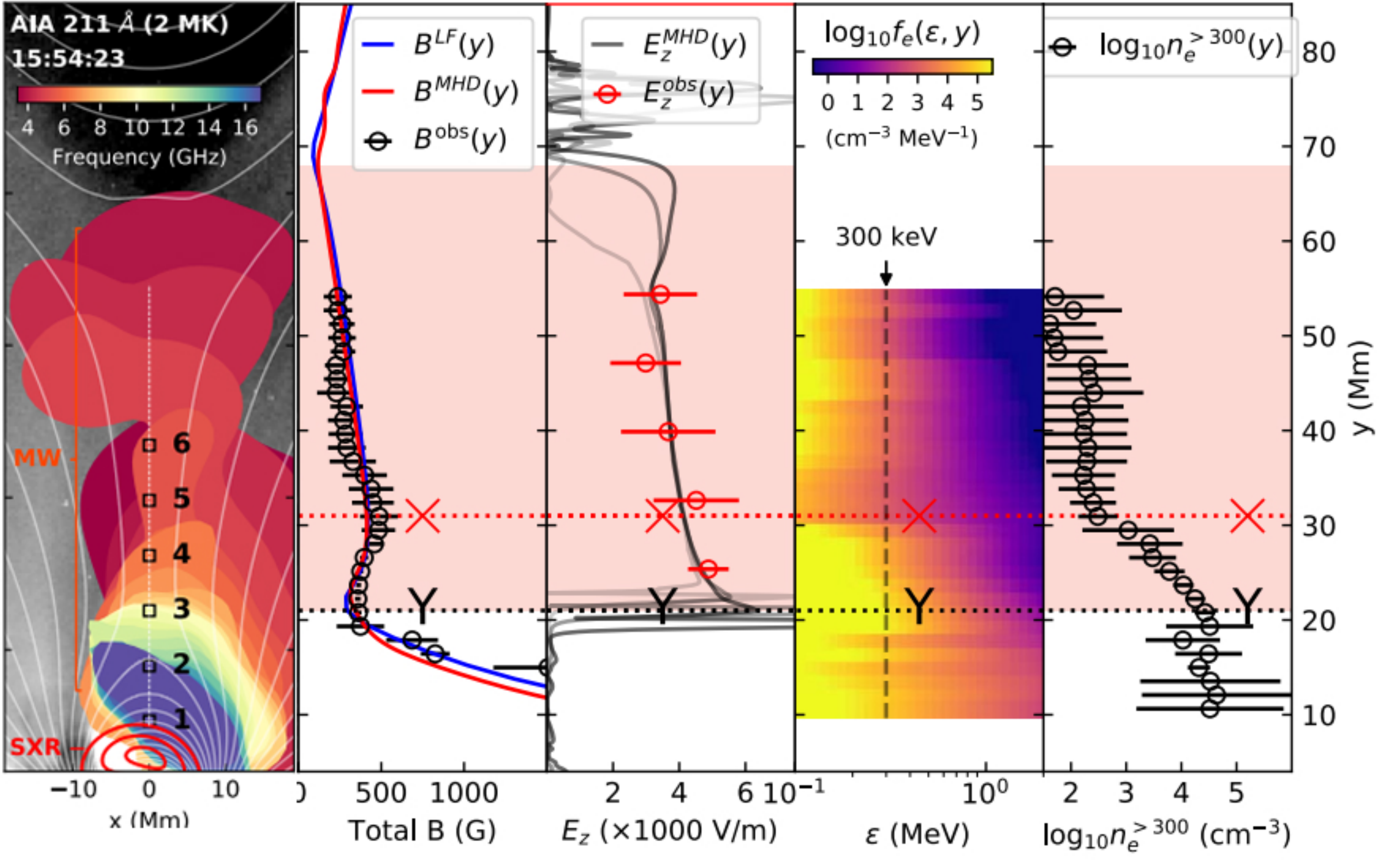}}
{\caption{Summary of results from \cite{2020NatAs...4.1140C} showing the EOVSA radio emission along the plasma sheet of the 2017 Sep 10 flare, the measured magnetic field strength showing a peak at the X point and dip at the cusp region (Y), the electric field, the electron energy spectra, and the number of nonthermal electron.}\label{fig:Chen}}
\end{figure}

Outstanding problems in the physics of flares include those of magnetic energy release \citep{2014GeoRL..41.3710D}, particle acceleration, and particle transport. As flare energy release requires the participation of a large coronal volume sometimes comparable to the size of the entire active region \citep{2011LRSP....8....6S}, one of the key challenges lies in the lack of measurements for key physical parameters in a broad region around the flare energy release site. At centimeter wavelengths, gyrosynchrotron emission – radiation from nonthermal electrons with energies of 10s of keV to several MeV gyrating in a magnetic field – illuminates any magnetic coronal loop to which energetic electrons have access, showing when and where accelerated electrons are present. Inversion of the gyrosynchrotron spectrum allows both the magnetic field in the flaring source and the electron energy distribution to be deduced, as well as their spatiotemporal evolution. EOVSA has recently demonstrated the power and utility of broadband (2.5–18 GHz) imaging spectropolarimetry with observations of the celebrated flares of September 2017 \citep[e.g.][]{2018ApJ...852L...9S,2020NatAs...4.1140C}. EOVSA measures both the dynamically changing magnetic field strength \citep{2020Sci...367..278F} and connectivity \citep{2020ApJ...895L..50C} of the erupting source (e.g. Figure~\ref{fig:Chen}), otherwise invisible at other wavelengths, and the spatially and temporally evolving electron distribution function \citep{2022Natur.606..674F}, placing new constraints on the magnetic energy release in the flaring source and on electron acceleration \citep[see the white papers by][]{Chen2022a,Gary2022b}. 

However, the limited dynamic range brought about by EOVSA’s sparse array of 13 antennas (typically less than 100:1) prevents the precise measurements of the microwave spectra outside the immediate site of the dominant microwave emission. Higher dynamic range and image fidelity are also required for investigating such important phenomena as non-powerlaw, broken powerlaw, and pitch angle dependence of the charged particle distribution, which have subtle spectral effects not discernible with EOVSA. FASR’s superior imaging in this frequency range (2–20 GHz) is expected to improve on the image dynamic range of EOVSA by more than two orders of magnitude (up to 10000:1), offering a revolutionary view of flare energy release and particle acceleration. Likewise, the extended frequency range of FASR towards lower frequencies (0.2–2 GHz) would allow tracking the charged particles into and above the erupting magnetic flux rope \citep{2020ApJ...895L..50C}, and hence would allow measurements of similar physical parameters in these largely unexplored regions in the flux rope and around the coronal shock associated with the erupting flux rope.  It is here that the interplay between flare- and shock-driven particle acceleration is suspected to seed solar energetic particles \citep[SEPs;][]{2016ApJ...830...28P} so important for their space weather effects. 

FASR’s lower frequency band (0.2–2 GHz) also provides access to plasma diagnostics to higher heights than are unavailable at higher frequencies. For example, coherent spike bursts and type III bursts driven by electron beams accompany energetic phenomena such as solar flares \citep{1997ApJ...480..825A} and jets \citep{2012ApJ...754....9G}. They are intimately related to the energy release process \citep{1998ARA&A..36..131B,2002A&A...383..678B} associated with magnetic reconnection in both the impulsive phase of flares and in post-flare loops. The high frequency and time resolution of the JVLA has been used to demonstrate that the trajectories of electron beams can be traced from the magnetic reconnection site in X-ray jets and flares \citep{2013ApJ...763L..21C,2018ApJ...866...62C}. In addition, reconnection-driven termination shock fronts can be mapped with the multi-frequency centroiding technique \citep{2015Sci...350.1238C,2021ApJ...911....4L}, which is also applicable to mapping CME-driven shocks from low to middle corona (\citealt{2019NatAs...3..452M}; see white paper by \citealt{Chen2022b}). 

\subsection{The Drivers of Space Weather}

The term ``space weather” refers to phenomena that disturb the interplanetary medium and/or affect the Earth and near-Earth environment \citep{2004ASSL..314....1L}. This includes recurrent structures in the solar wind (fast solar wind streams, co-rotating interaction regions), the ionizing radiation and hard particle radiations from flares, coronal mass ejections, shock-accelerated particles, and the radio noise itself. Each of these can cause phenomena on Earth that have societal impacts. The science of space weather is in a fledgling state but its relevance to modern society is highlighted by the OSTP’s National Space Weather Strategy and Action Plan (2019), and subsequent passage of the PROSWIFT Act (https://www.congress.gov/bill/116th-congress/senate-bill/881). Moreover, its relevance as a touchstone for exo-space weather and planet habitability is increasingly being recognized, as well as to human deep space exploration. Comprehensive and diverse observations are needed to understand space weather as a complex system. FASR will be sensitive to eruptive flares, CMEs, filament/prominence eruptions, and type II (shock-driven) radio bursts (see white paper by \citealt{Chen2022c}). 
With its unique ability to perform broadband imaging spectropolarimetry, FASR will be able to simultaneously image the basic shock driver (flare or CME signature), the response of the atmosphere to the driver \citep[chromospheric/coronal waves and coronal dimming;][]{2005ApJ...620L..63W}, and the shocks themselves as a system. In doing so FASR will yield unique new insights into the relationship between flares, CMEs, SEPs and their space weather impacts. See white papers by \citet{Saint-Hilaire2022} and \citet{White2022b}.

\subsection{Energetics of the Quiet Solar Atmosphere}

The heating of the solar chromosphere and corona remains an enduring problem. Radio emission from the quiet Sun atmosphere is dominated by thermal free-free emission but nonthermal processes also play a role \citep[e.g.][]{2018ApJ...852...69S}. At its highest frequency, where its spatial resolution is also highest, FASR will image the middle chromosphere. 
By systematically tuning to lower frequencies, the height from which the (optically thick) emission originates moves from the top of the chromosphere into the corona (see white paper by \citealt{Kobelski2022}), again providing a 3D probe into the temperature, density, and the magnetic field structure \citep{2019SoPh..294....7R} of this mysterious interface region of the solar atmosphere. Resonant wave heating \citep[e.g.][]{2011ApJ...736....3V} represents an important class of models for coronal heating and makes specific predictions of the location and time scales of energy deposition. Using free-free diagnostics, FASR observations of the resulting temperature changes in this region encompassing both the chromosphere and the adjacent corona can address the validity of such predictions. The role of “nanoflares” \citep{2017NatAs...1..771I, 2017ApJ...844..132W} in the energetics of the lower atmosphere is also under active investigation. The energetics of such small events depend critically on their thermal properties and the role of accelerated electrons \citep{2018ApJ...852...69S,2020ApJ...895L..39M}. The latter, which constitutes the ``smoking gun'' of the nanoflare-driven coronal heating scenario, may yield detectable radio signatures against the background by a highly sensitive, high-imaging quality instrument like FASR (see white paper by \citealt{Mondal2022}). Again, the combination of FASR’s frequency coverage and imaging can apply diagnostics to quantitatively address both thermal/nonthermal contributions of such events to the energy budget. Another quiet Sun FASR science use case is to constrain the precursor environment of flares (active region filaments) and CMEs (e.g., coronal cavities). 

\section{Technical Overview}

The FASR concept presented here differs in some respects from previous versions. First, significant investments have been made in ground-based low-frequency arrays ($<$0.2 GHz): LOFAR, MWA, LWA, OVRO-LWA. A separate antenna subsystem (FASR-C) to observe frequencies $<$0.2 GHz is redundant because the OVRO-LWA will support solar-dedicated science in this frequency band simultaneously with FASR. Thus, the notional plan described here is focused on the FASR-A and FASR-B arrays. Second, recent technological advances, along with advancing science imperatives, have made it important and achievable to increase the number of antennas, for better image quality and dynamic range, and to increase the instantaneous bandwidth by a factor of 8 to improve the time cadence to 0.1~s. The increase in data volume implied by these changes is no longer considered an impediment.  The site for FASR has not been decided. We have discussed sharing a site in Nevada proposed for a new astrophysics array called DSA-2000 \citep{2022MNRAS.514.2614C}, should that proposal be successful, to constrain infrastructure costs.  The previous costing exercise put the cost of FASR at approximately \$58M (FY2010). FASR has always been conceived as a midscale project that could be implemented quickly and, if need be, in phases.  By utilizing technological advances and economies of scale for the production of a larger number of antennas, we expect the total cost to remain well within the mid-scale range.


\section{Instrument Specifications}

The science objectives outlined in §3 flow down to the high-level science and technical specifications summarized in Table 1. FASR is designed to perform Fourier synthesis imaging using well-established interferometric techniques. For an array of $N$ antennas there are $\sim N^2/2$ independent antenna pairs, each of which measures a single Fourier component, or complex visibility, of the Sun’s radio brightness distribution at a given frequency, time, and spatial scale. The ensemble of antenna pairs, or antenna baselines, therefore measures many Fourier components. Fourier inversion of the visibility measurements yields an image of the Sun’s radio brightness at a given frequency, time, and polarization. Deconvolution techniques are then used to remove the effects of the point spread function, the response of the instrument to a point source.

\begin{wrapfigure}{r}{0.6\textwidth}
    Table 1: FASR Specifications\\
    \includegraphics[width=1\textwidth]{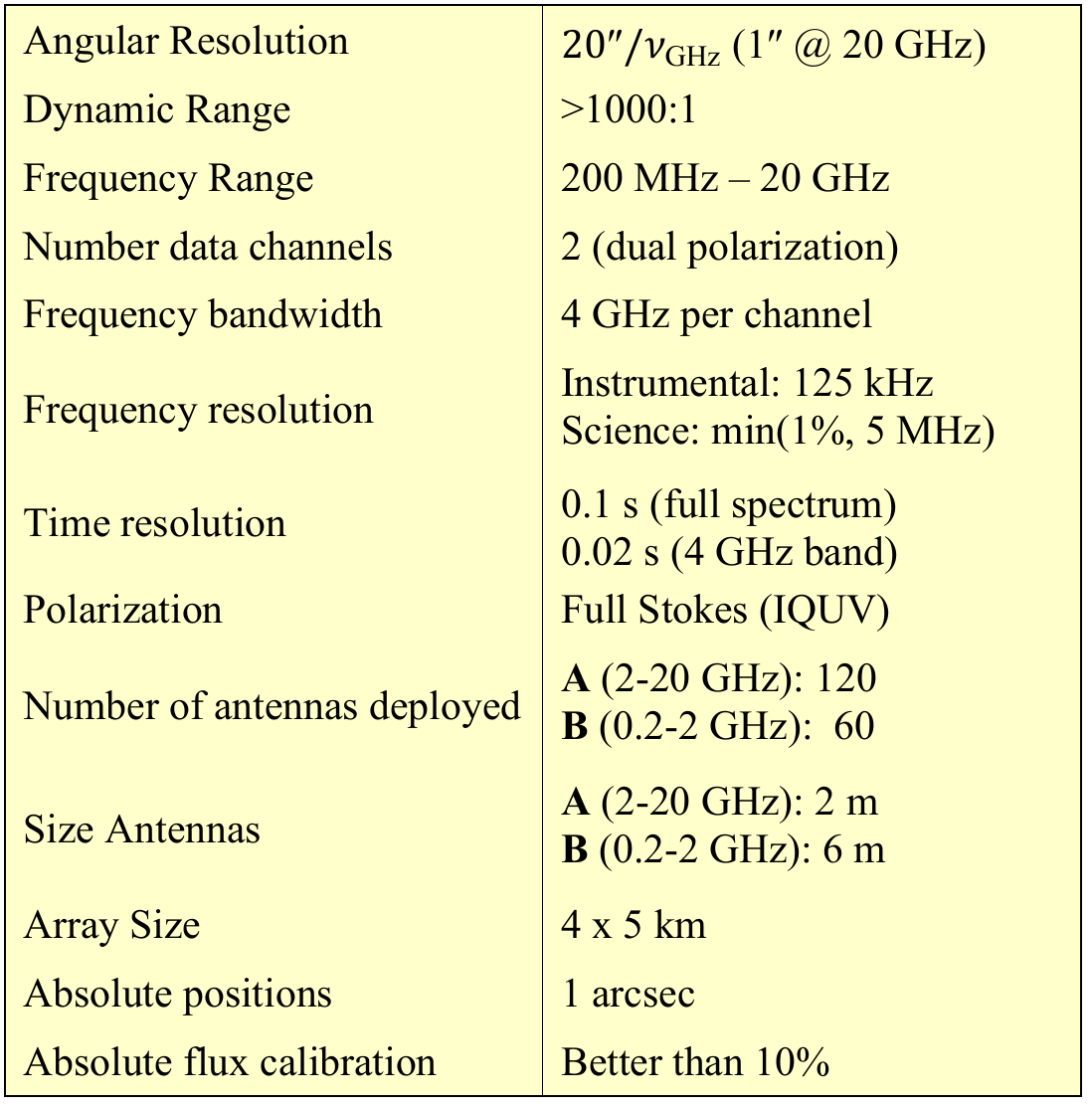}
\end{wrapfigure}

FASR observations will cover an unprecedented two-orders-of-magnitude frequency range, 0.2 to 20~GHz (wavelengths from 1.5~cm to 1.5~m) using two separate arrays of antennas, denoted FASR A and FASR B. Each array provides frequency coverage over roughly a decade of bandwidth: $\sim$2-20 GHz (FASR A); $\sim$0.2-2 GHz (FASR B). The number, type, and configuration of the antennas in each array are chosen to address the key science objectives summarized in §3. See the SSP white paper by \cite{Chen2022a} for an example configuration for FASR A. The antennas will be distributed over an area with a diameter $\sim$4~km, providing an angular resolution of 1$\arcsec$ at 20~GHz. The angular resolution scales linearly with wavelength from this fiducial value and is sufficient to resolve typical solar radio sources as well as being well matched to the capabilities of existing and planned space missions. The antenna size for each sub-array is determined by the competing needs of full-disk imaging on the one hand, and sufficient sensitivity to calibrate the instrument against sidereal sources on the other.

\section{Technology Drivers}

FASR exploits radio-astronomy techniques that have a substantial heritage, namely, Fourier synthesis imaging and image deconvolution. Moreover, a FASR pathfinder array has already been deployed (EOVSA) that leveraged the previous FASR design effort and has demonstrated the fast-tuning technology as well as the approaches taken to calibration and imaging, thereby retiring technical elements that were previously regarded as carrying minor to moderate risk. From a technical standpoint, there are no ``tall poles” barring the way or significant risk elements. The FASR project will exploit the latest technological advances in broadband feeds, broadband signal transmission, high-speed digital processing, high-speed digital networks, data storage, and computer processing.  we expect that continued commercial advances will improve on cost and schedule (making everything cheaper and easier over time), but we can build FASR with current technology. There is no technological impediment to building FASR today.

\section{Project Organization, Status, and Schedule}

The FASR project is conceived as a consortium of university partners (NJIT, UC/Berkeley, Caltech, U Michigan, and others) and national observatories (NRAO and, ideally, NSO). The FASR pathfinder, EOVSA, was completed in 2017 and is engaged in building the solar radio community within the U.S. There is a sizeable hard X-ray community familiar with Fourier synthesis imaging of flares with the now-decommissioned RHESSI hard X-ray imager, as well as a growing community of ALMA users for studies of the chromosphere. We expect this growth to continue through the current solar cycle with the availability of EOVSA and other imaging arrays (OVRO-LWA, JVLA, LOFAR, MWA).
Having demonstrated the science with EOVSA, the FASR project is now undergoing a rebuild, kicked off by a workshop in December 2021 (https://fasr21.ssl.berkeley.edu/), and with a top-level design review scheduled for January 2023 that will lead to a refined concept to be presented to the Solar and Space Physics decadal survey Solar and Heliospheric Panel when they meet in 2023. A renewed decadal endorsement in 2024 would put the project in position for the submission of an MSRI-R1 design study proposal in 2025, which is needed to expand the team and bring the project to a state of readiness for an MSRI-R2 construction proposal two years later (2027). Due to the technical feasibility of the project, actual construction can be completed in 3-4 years (by 2032).


\section{Cost Estimate}

The most comprehensive costing exercise for FASR was conducted for a proposal to the NSF Atmospheric Sciences Mid-Sized Infrastructure (MSI) Opportunity a decade ago---a program that was subsequently canceled---for which detailed costs were developed for a reference instrument. The proposal was prepared by the FASR team, with task leaders providing detailed cost estimates for each element of the work breakdown structure. Both the budget and schedule were reviewed by an external ``red team”. This subsequently served as the basis for the cost estimate of \$58M, which after adding an 18\% contingency resulted in a total cost of \$68M conveyed to Astro2010.  They evaluated the project using the CATE process (analogous to the TRACE process) and returned a very conservative $\sim$\$100M cost estimate. In addition, costs were analyzed for operations and maintenance of FASR, including a data management plan. These are no longer relevant.

One of the main tasks for the upcoming FASR design review in January 2023 will be to make our best estimate of the current cost of FASR, to be proposed for further study in 2025.  We have not yet gone through the required exercise, but we fully expect that the cost of an instrument meeting the specifications shown in Table 1 will be no more than the \$70 M estimated previously. This is largely due to less expensive sources for antennas and the vast improvement in bandwidth of both analog and digital hardware in the past 10 years. Although we cannot provide a cost breakdown at this time, we expect to have this information in time to present it to the Solar and Space Physics decadal survey Solar and Heliospheric panel when they meet in 2023.


%
\newpage
\bibliography{FASR_SSP.bib,All_WP.bib}{}
\bibliographystyle{aasjournal_bc}

\end{document}